\documentclass[english]{article}
\usepackage[T1]{fontenc}
\usepackage[latin9]{inputenc}
\usepackage{amsmath}
\usepackage{amssymb}
\usepackage{graphicx}
\usepackage{esint}
\usepackage{babel}
\begin{document}

\title{Anelasticity in Flexure Strips Revisited}

\author{Clive C. Speake}

\maketitle
School of Physics and Astronomy, University of Birmingham, Edgbaston,
Birmingham B15 2TT, UK
\begin{abstract}
This note reviews previous analyses by the author of the damping produced
by the anelasticity of a simple flexure element that is loaded in
tension by an extended object such as a beam balance. The correct
calculation of the anelasticity of a simple flexure appeared in an
appendix in Quinn et al (1995)\cite{Quinn_et_al_1995} where the change
in the gravitational potential energy due to the shortening of the
flexure was calculated enabling expressions for the elastic energy
and its associated losses to be derived. Publications prior to this
paper did not include this lossless term which led to incorrect predictions
of the anelastic losses in flexure pivots in Quinn et al (1987) \cite{phil_mag_anelasticity}.
In this current paper the derivation of the result is given in such
a way that it can be easily contrasted with the expressions in these
earlier papers. I also extend the methodology to calculate the elastic
and gravitational energy associated with the motion of a suspended
object whose dimensions are significantly smaller than the length
of the flexure.
\end{abstract}

\section*{PACS: }

07:10.Lw Balance Systems, 62.20.D Elasticity, 04.80.Nn Gravitational
Wave detectors

\section{Introduction}

Many physical measurements are derived from mechanical oscillators.
For example, the measurement of mass relies on the precision achievable
in common balances and thermal noise in mechanical suspensions plays
a key role in the sensitivity of gravitational wave detectors. It
has been known for some time that anelasticity in material suspensions
produces non-linearity and is also a source of thermal noise (\cite{phil_mag_anelasticity},\cite{Erice_noise_paper},\cite{Saulsen}).
Derivation of the key equations describing the elastic behaviour of
a simple flexure-strip supporting a beam balance has been given in
previous papers \cite{phil_mag_anelasticity,Speake_Fund_limits_beam_balance,Quinn_et_al_1987,quinn_precise_weighing}.
However, looking over these papers, that were published now more than
20 years ago, it appears appropriate to collect the key results together
and present them in a coherent way. In particular a treatment of anelasticity
that included the change in gravitation potential energy due to the
shortening of the bending flexure was given in an appendix in reference
\cite{Quinn_et_al_1995}. In this current paper I will revisit the
results of these earlier papers and present them in the context of
the work reported in the 1995 paper to provide confirmation of the
correct result. I will also use the same methodology as used in reference
\cite{Quinn_et_al_1995}to derive expressions for the anelastic losses
in a simple pendulum suspension.

\section{Key Equations}

I will derive the general equations governing the quasi-static stiffness
of a simple flexure element of uniform cross section. As shown in
Figure 1, a flexure-strip, of second moment of area $H$ and length
$L$, supports a load of weight $W$. We will assume that the flexure
has a rectangular cross-section with width $b$ and thickness $t$
and in this case we have $H=\frac{bt^{3}}{12}$. A torque, $\tau$,
and a horizontal force, $F$, are applied to the free end of the flexure
at $x=L$. This results in a reaction force $-F$ and a reaction torque
$\tau_{0}$ at the upper end of the flexure. As described in reference
\cite{Quinn_et_al_1987} and basic undergraduate texts, provided that
the radius of curvature of the flexure is much larger than its thickness,
Hooke's law can be used to equate the moment of the forces, acting
at any point, to the curvature of the flexure, which in turn can be
found in terms of the stress distribution across its cross-section.
This method is that adopted by previous authors (see references given
in \cite{Quinn_et_al_1987}), however in reference \cite{Quinn_et_al_1987},
we discuss limitations to this method such as those imposed due to
the finite width of the flexure and its Poisson's ratio. The equation
describing the bending of the flexural element is given, more generally,
by a fourth order equation but, here, we ignore any shear forces due
to a distributed load (\textit{ie} the mass per unit length of the
flexure is considered to be negligible). Thus our approach differs
from the analyses of other authors: in reference \cite{Piergiovanni},
for example, the problem of the dynamic excitation of the suspension
elements for mirrors for gravitational wave detectors is analysed
and the mass per unit length of the suspensions is key to the analysis.

We therefore seek solutions to the second order differential equation
specifying the bending moment as a function of position that can be
written in terms of the torque, $\tau$, acting at the free end and
the applied forces, 
\begin{equation}
M\left(x\right)=EH\frac{d^{2}y}{dx^{2}}=\tau+F\left(L-x\right)-W\left(y(L)-y(x)\right)\label{eq:1-1}
\end{equation}
where $E$ is the Young's modulus of the flexure. A solution to this
problem, that satisfies the boundary conditions given by the positions
and tangents at both ends of the flexure is,
\begin{equation}
y=\frac{F}{\alpha W}\left(\tanh\alpha L\left(\cosh\alpha x-1\right)+\alpha x-\sinh\alpha x\right)+\frac{\tau}{W\cosh\alpha L}\left(\cosh\alpha x-1\right)\label{eq:1b}
\end{equation}
where $\alpha^{2}=\frac{W}{EH}$. Plots of equation \ref{eq:1b}for
the cases where there are either torques or forces applied to the
flexure are shown in Figures 2a and 2b where the physical values of
the parameters that characterise the flexure are those of the flexure
described in \cite{Quinn_et_al_1987}. This solution can then be used
to express the tangent angle, $\theta_{0}$ and transverse displacement,
$y_{0}$, as shown in Figure 1, in terms of the applied torque and
force. It is convenient to write these relations in the form a compliance
matrix, $\underline{\underline{C}}$,
\begin{equation}
\left(\begin{array}{c}
y_{0}\\
\theta_{0}
\end{array}\right)=\underline{\underline{C}}\left(\begin{array}{c}
F\\
\tau
\end{array}\right),\label{eq:2}
\end{equation}
with
\begin{equation}
\underline{\underline{C}}=\left(\begin{array}{cc}
C_{11} & C_{12}\\
C_{21} & C_{22}
\end{array}\right),\label{eq:3a}
\end{equation}
where 
\begin{equation}
C_{11}=\frac{\alpha L\cosh\alpha L-\sinh\alpha L}{\alpha W\cosh\alpha L},\label{eq:3b}
\end{equation}
\begin{equation}
C_{12}=C_{21}=\frac{\cosh\alpha L-1}{W\cosh\alpha L},\label{eq:3c}
\end{equation}
and
\begin{equation}
C_{22}=\frac{\alpha}{W}\tanh\alpha L.\label{eq:3d}
\end{equation}

\begin{center}

\includegraphics[width=10cm]{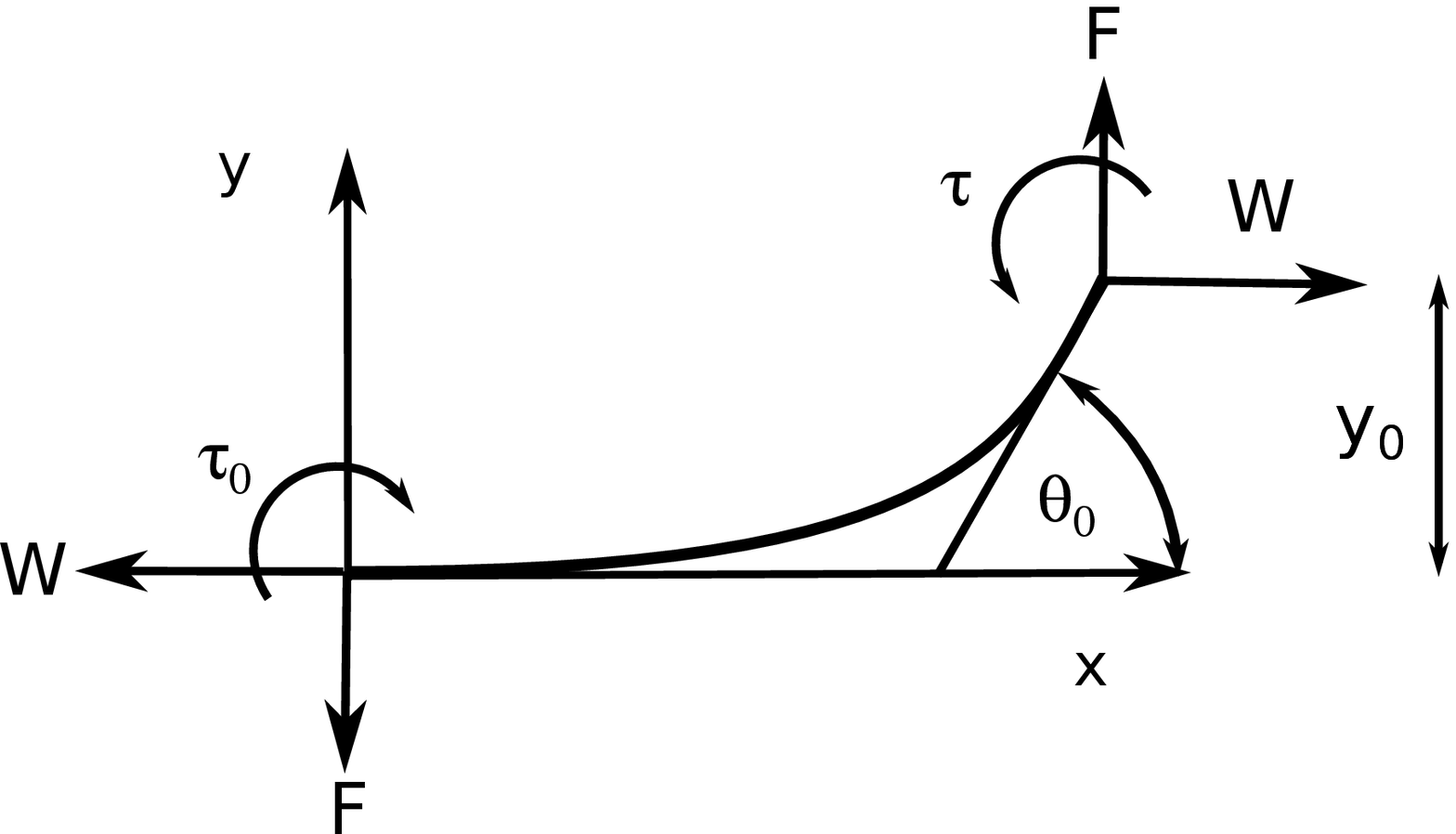}

\end{center}

Figure 1 A schematic the flexure element. The flexure is loaded with
force, $W$, due to the weight of the suspended object, and torques
due to a horizontal force, $F$, and a torque, $\tau$. This results
in a transverse displacement $y_{0}$ and and an angular deflection
$\theta_{0}$ at the free end of the flexure. Note that Earth's gravity
is vector points in the positive $x$ direction.

\medskip{}

\begin{center}

\includegraphics[width=10cm]{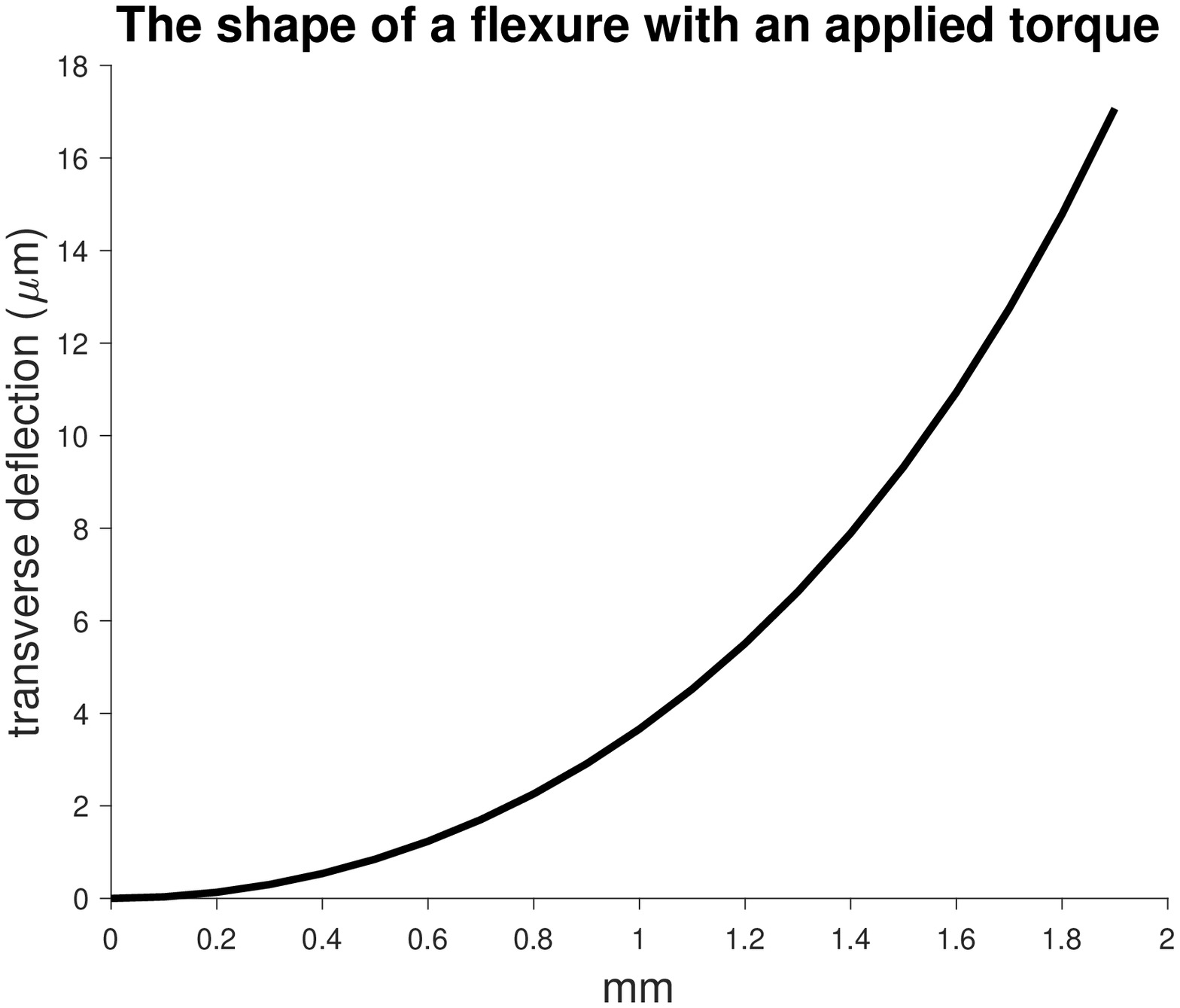}

\end{center}

Figure 2a 

\medskip{}

\begin{center}

\includegraphics[width=10cm]{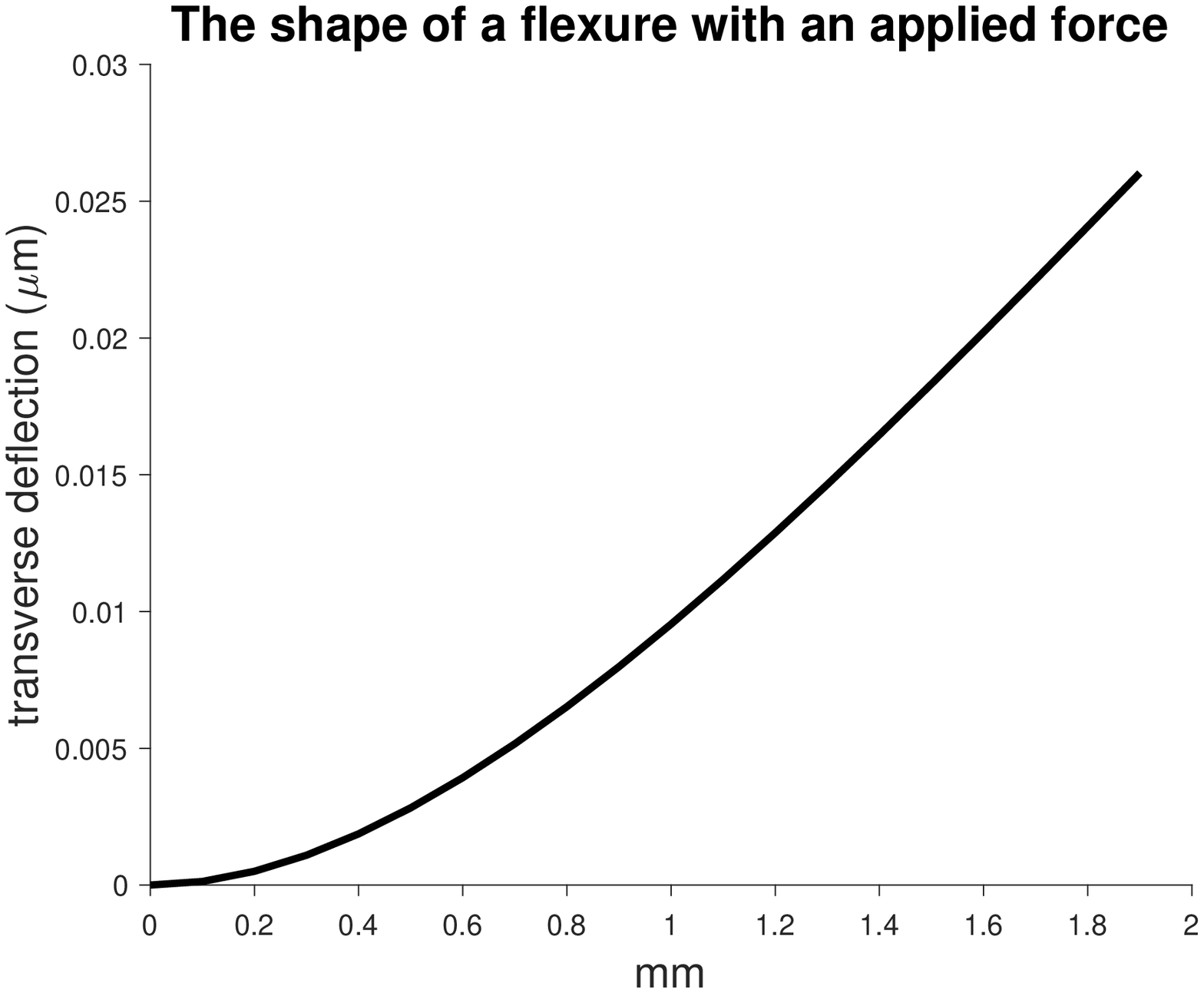}

\end{center}

Figure 2b. 

Figures 2a and 2b show the shape of the flexure as given by equation
\ref{eq:1b} for the cases where a simple torque or force is applied
respectively. In Figure 2a the torque has an abitrary magnitude of
$10^{-3}Nm$ and in Figure 2b the force applied has an arbitrary magnitude
of $10^{-3}N$. The flexure dimensions are given as $t=50\mu m$,
$b=24mm$, $L=20mm$ with $E=130GPa$ and $W=40N$.

\medskip{}

In our previous analyses we have ignored the force\textbf{, $F$,}
and the omission of the applied force is an approximation that can
be understood as follows. The suspended object is assumed to be oscillating
in Figure 1 and has arrived at the deflection shown there by moving
through angle $\theta_{0}$. The force acting on the end of the flexure
due to this motion is the inertial force proportional to the object's
linear acceleration. On the other hand, the torque, $\tau$, is due
to the angular inertial acceleration and is proportional to the objects
moment of inertia. In the limit that the object's radius of gyration
is much larger than the distance of its centre of mass from the centre
of rotation (given as $R_{f}$ in the discussion below) it is reasonable
to assume that the suspended object applies a pure torque to the end
of the flexure and that the force is negligible. When an extended
object such as a beam-balance is attached to the flexure, as was the
case in our previous work, we can consider that a pure torque is being
applied, on the other hand when the suspended object is a point mass
we can assume that only a force acts. It is worth noting that a complete
treatment of the dynamical equations of a simple flexure that is supporting
a mass of a given moment of inertia has been analysed by Haag \cite{Haag}.
In this analysis it is shown that the inertial stability of the oscillating
bob can change the shape, effective rotation axis and the energy stored
in the flexure. This full analysis (that has to be treated numerically)
is not necessary for quasi-dc applications as is the case here. 

In what follows we will use an superscript {*} to indicate a result
that is revised in Section III. The well-known result (see for example
\cite{Quinn_et_al_1987}) is that the bending stiffness, $\kappa_{t}$,
is
\begin{equation}
\kappa_{t}=\frac{\tau}{\theta_{0}}=\frac{1}{C_{22}}=\frac{W}{\alpha}\coth\alpha L,\label{eq:4}
\end{equation}
This result follows from equation \ref{eq:2} by setting $F=0$. It
is evident that the stiffness of the flexure does not depend on its
length in the limit that $\alpha L>>1$ and for this reason, in reference
\cite{Quinn_et_al_1987}, the product $\alpha L$ was chosen to be
approximately two. It is helpful to keep the length of flexure as
small as possible as this increases the stiffness of higher order
modes and reduces their coupling to the simple rotational mode that
we are modelling. It is tempting to consider that the trajectory of
the supported object is determined by an effective radius that is
determined by the ratio of transverse deflection, $y_{0}$, and the
angular deflection, $\theta_{0}$. We could define
\begin{equation}
r_{f}^{*}=\frac{y_{0}}{\theta_{0}}=\frac{C_{12}}{C_{22}}=\alpha^{-1}\tanh\left(\frac{\alpha L}{2}\right).\label{eq:5a}
\end{equation}
Further we could then assume that the system behaves as a compound
pendulum and possesses a gravitational potential energy determined
by this length. A gravitational stiffness, $\kappa_{g}^{*}$, could
then be defined as
\begin{equation}
\kappa_{g}^{*}=Wr_{f}^{*}=\frac{W}{\alpha}\tanh\left(\frac{\alpha L}{2}\right)\label{eq:5b}
\end{equation}

The total stiffness, $\kappa_{t}$, then comprises the gravitational
term, $\kappa_{g}^{*}$, and another term that could be considered
to be the 'intrinsic' elastic stiffness of the flexure, $\kappa_{el}^{*}$,
where

\begin{equation}
\kappa_{t}=\kappa_{el}^{*}+\kappa_{g}^{*}.\label{eq:6a-1}
\end{equation}
Hence we find

\begin{equation}
\kappa_{el}^{*}=\frac{W}{\alpha}csh(\alpha L).\label{eq:6c}
\end{equation}
 This type of analysis has also been used to derive expressions for
the stiffnesses of end suspensions of beam balances in reference \cite{Quinn_et_al_1987}.
We will revise these equations in Section III below. 

It is generally accepted that a significant contribution to the damping
of the motion of flexures comes from material anelasticity. Of particular
importance to the field of mass metrology is the anelastic behaviour
of a beam balance when it is subjected to a change in stress. This
is referred to as the anelastic after-effect. Anelasticity, although
observed by Henry Cavendish \cite{Cavendish}, is analysed in detail
in connection with measurements of weak forces in reference \cite{phil_mag_anelasticity}.
Following reference \cite{phil_mag_anelasticity}, the stiffness can
be expanded in terms of a frequency dependent modulus defect $\delta E\left(\omega\right)$,
\begin{equation}
\kappa_{el}\left(E\right)=\kappa_{el}(E_{0}+\delta E\left(\omega\right))\backsimeq\kappa_{el}\left(E_{0}\right)+\frac{d\kappa_{el}}{dE}\delta E\left(\omega\right),\label{eq:7a-1}
\end{equation}
where $E_{0}$ is the component of the Young's modulus that obey's
Hooke's law. It is also useful to define
\begin{equation}
\kappa_{anel}=\left[-\frac{\alpha}{2}\frac{d\kappa_{el}}{d\alpha}\right]\frac{\delta E\left(\omega\right)}{E_{0}}.\label{eq:7b-2}
\end{equation}
The frequency dependent part of the modulus defect, $\delta E\left(\omega\right)$,
has a real and imaginary part. The latter, $\delta E^{i}\left(\omega\right)$,
is responsible for damping. In reference \cite{phil_mag_anelasticity}
we were only interested in the damping and the anelastic after-effect
and thus we ignored the real part of the modulus defect. We assumed
that the total modulus defect comprised the sum of a distribution
of relaxation processes that all have the same relaxation strength,
$\delta e$, where
\begin{equation}
\delta E^{i}\left(\omega\right)=\delta e\intop_{\tau_{0}}^{\tau_{\infty}}f\left(\tau\right)\frac{\omega\tau}{1+\omega^{2}\tau^{2}}d\tau.\label{eq:7b}
\end{equation}
The weighting factor, $f\left(\tau\right)$, of the dissipation mechanisms
of relaxation time $\tau$ (not to be confused here with the applied
torque) is normalised as follows,
\begin{equation}
\intop_{\tau_{0}}^{\tau_{\infty}}f\left(\tau\right)d\tau=1.\label{eq:7c-1}
\end{equation}
Our measurements, that extended to oscillation periods of 690 $s$,
indicated that the imaginary component of the modulus was independent
of frequency even at these low frequencies. This behaviour is reproduced
with
\begin{equation}
f\left(\tau\right)=\frac{1}{\ln\left(\tau_{\infty}/\tau_{0}\right)}\frac{1}{\tau},\label{eq:7d}
\end{equation}
and for $\omega\tau_{0}\ll1$ and $\omega\tau_{\infty}\gg1$. Thus
by performing the integrations (see \cite{phil_mag_anelasticity})
we can define 
\begin{equation}
\delta E^{i}\left(\omega\right)=\delta e\frac{\pi}{2\ln\left(\tau_{\infty}/\tau_{0}\right)}.\label{eq:7e-1}
\end{equation}
Using \ref{eq:6c}, we find
\begin{equation}
\left[-\frac{\alpha}{2}\frac{d\kappa_{el}}{d\alpha}\right]=\frac{W}{2\alpha}\frac{1}{\sinh\alpha L}\left(1+\alpha L\coth\alpha L\right).\label{eq:7b-1}
\end{equation}
We can find the variation of $\kappa_{anel}^{*}$ with the change
in length, $L$, by expanding \ref{eq:7b-1} around $L\approx\alpha^{-1}$,
for constant load and flexure geometry. We find
\begin{equation}
\kappa_{anel}^{*}=i\frac{\delta E^{i}}{E_{0}}\frac{W}{\alpha}\left(0.98-1.60(\alpha L-1)+O\left(\alpha L-1\right)^{2}\right).\label{eq:7e}
\end{equation}
We also can show that $\kappa_{anel}^{*}$ tends to zero as $L$ tends
to infinity. This was a result that was revised in the later publication
\cite{Quinn_et_al_1995}and is discussed further below.

See references \cite{phil_mag_anelasticity} and \cite{expt_and_theory_anelasticity}
for more details of the calculation of the anelastic-after-effect
and the damping produced by anelasticity and references to the work
of other authors. 

\section{Revision of these equations}

Section II derives the results that were developed in our papers up
until reference \cite{Quinn_et_al_1995} in 1995. The result of equations
\ref{eq:6c} and \ref{eq:7e} is that the anelastic stiffness reduces
exponentially as the flexure becomes longer or thinner. It makes sense
physically that the damping reduces to zero as the second moment are
area reduces to zero. We do not, however, expect that it reduces by
simply making the flexure-strip longer as the bending moment at the
end of the flexure cannot physically reduce to zero as the flexure
becomes longer. This is consistent with the stiffness of the flexure
becoming independent of the length of the flexure in the limit that
$L\gg\alpha^{-1}$. The calculations given above were also not consistent
with experimental results obtained in \cite{Quinn_et_al_1995}. At
this point I tried another approach where the elastic stored energy
and the gravitational energy were treated separately\textbf{ }and,
importantly the change in gravitational potential energy was calculated
as being due\textbf{ }to the change in height of the load, $W,$ as
the flexure deforms. The stored elastic energy can be written (see
Appendix A in \cite{Quinn_et_al_1995}),
\begin{equation}
V_{el}=\frac{1}{2EH}\intop_{0}^{L}M^{2}\left(x\right)dx,\label{eq:8a}
\end{equation}
where again $M\left(x\right)$ is the bending moment:
\begin{equation}
M\left(x\right)=EH\frac{d^{2}y}{dx^{2}}.\label{eq:8b}
\end{equation}
If we confine ourselves to the situation where $F$ is insignificant
in equation \ref{eq:2} then from equation \ref{eq:1b} and \ref{eq:3d},
\begin{equation}
y(x)=\frac{\theta_{0}}{\alpha}\frac{\left(\cosh\left(\alpha x\right)-1\right)}{\sinh\left(\alpha L\right)}.\label{eq:8c}
\end{equation}
Thus we can show that 
\begin{equation}
V_{el}=\frac{1}{2}\frac{W}{\alpha}\theta^{2}\frac{1}{2}\left(\coth\left(\alpha L\right)+\frac{\alpha L}{\sinh^{2}\left(\alpha L\right)}\right),\label{eq:8d}
\end{equation}
where we have deliberately separated out the two factors of one half.
We can define an elastic stiffness that comes from this approach as
\begin{equation}
\kappa{}_{el}=\frac{W}{2\alpha}\left(\coth\left(\alpha L\right)+\frac{\alpha L}{\sinh^{2}\left(\alpha L\right)}\right).\label{eq:8e}
\end{equation}
The bending of the flexure results in its length projected onto the
vertical direction being shortened. To second order in the bending
angle, the change in vertical height of the suspended load can be
calculated to be 
\begin{equation}
\varDelta L=\frac{1}{2}\intop_{0}^{L}\left(\frac{dy\left(x\right)}{dx}\right)^{2}dx.\label{eq:9a}
\end{equation}
This results in a change in the gravitational energy of
\begin{equation}
V_{g}=W\varDelta L=\frac{1}{2}\frac{W}{\alpha}\theta^{2}\frac{1}{2}\left(\coth\left(\alpha L\right)-\frac{\alpha L}{\sinh^{2}\left(\alpha L\right)}\right),\label{eq:9b}
\end{equation}
with an associated gravitational stiffness,
\begin{equation}
\kappa{}_{g}=\frac{W}{2\alpha}\left(\coth\left(\alpha L\right)-\frac{\alpha L}{\sinh^{2}\alpha L}\right).\label{eq:9c}
\end{equation}
We can define a radius of rotation as in the previous analysis in
Section 2,
\begin{equation}
r_{f}=\frac{1}{2\alpha}\left(\coth\alpha L-\frac{\alpha L}{\sinh^{2}\alpha L}\right).\label{eq:9d}
\end{equation}
We can see that 
\begin{equation}
V_{g}+V_{el}=\frac{1}{2}\kappa_{t}\theta^{2},\label{eq:10a}
\end{equation}
and
\begin{equation}
\kappa{}_{t}=\kappa{}_{g}+\kappa{}_{el}.\label{eq:10b}
\end{equation}
It is important to note that the equations from\ref{eq:8a} including
through to equation\ref{eq:10b}, where the gravitational energy term
is calculated in terms of the shortening of the flexure as it bends,
appeared in the appendix of \cite{Quinn_et_al_1995}. In this current
paper we are reiterating the importance of these results and explaining
further their interpretation.

These equations lead to the same result for the total stiffness as
in equation \ref{eq:4} but the elastic and gravitational stiffnesses
differ from the previous analysis, as given by equations \ref{eq:8d}
and \ref{eq:10b} compared with equations \ref{eq:6c} and \ref{eq:8d}.
However the geometrical interpretation of $r_{f}^{*}$ is a simple
function of the geometry of the stressed flexure and its use to estimate
the position of the effective centre of rotation of the compound pendulum
remains valid. This is an important parameter when designing a device
that has minimum sensitivity to horizontal ground vibrations and tilt.
We will therefore refer to $r_{f}^{*}$ as $R_{f}$, which we can
consider to be the location of the effective pivot axis, as measured
from the end of the flexure. The length $R_{f}$ is also the parameter
that enters into the calculation of the moment of inertia of the suspended
object about the effective point of rotation. The importance of $r_{f}$
lies in its use for calculating the gravitational potential energy
following equations \ref{eq:9c} and \ref{eq:9d}. We can calculate
the ratio $r_{f}$ to $r_{f}^{*}$ and express the result numerically
for the usual case when $L\approx\alpha^{-1}$,
\begin{equation}
\frac{r_{f}^{*}}{r_{f}}\approx1.57+0.13(\alpha L-1)+O\left((\alpha L-1)^{2}\right)...\label{eq:11}
\end{equation}
The previous analysis therefore overerestimates the value of $r_{f}$
by a factor of $1.57$. Therefore using $r_{f}^{*}$ to estimate the
gravitational potential energy of the system results in an error of
some $57\%$. 

Now that we have adequately accounted for the gravitational energy
involved in the flexure bending, as was already given in the 1995
publication, we can calculate the anelasticity from the stored elastic
energy. We can use equation \ref{eq:8d} to calculate the anelasticity
of the flexure in terms of the elastic stiffness. We find that 
\begin{equation}
\left[-\frac{\alpha}{2}\frac{d\kappa_{el}}{d\alpha}\right]=\frac{W}{4\alpha}\left(\coth\left(\alpha L\right)+\frac{\alpha L}{\sinh^{2}\alpha L}+\frac{2\left(\alpha L\right)^{2}\coth\alpha L}{\sinh^{2}\alpha L}\right).\label{eq:12}
\end{equation}
 We can present this result numerically in the case of $L\approx\alpha^{-1}$,
as follows,
\begin{equation}
\kappa_{anel}=i\frac{\delta E^{i}}{E}\frac{W}{\alpha}\left(0.98-1.04\left(\alpha L-1\right)+O(\alpha L-1)^{2}\right),\label{eq:13a}
\end{equation}
The anelastic component of the flexure stiffness behave\textbf{s}
in similar way to the expression given for $\kappa_{anal}^{*}$ in
the previous section for the case where $L\approx\alpha^{-1}$. However,
for $L\gg\alpha^{-1}$, the damping now tends to a value of 0.25 and
this expression for the anelastic stiffness is physically reasonable. 

\section*{4 Further calculations}

The results above are all relevant for the case where the suspended
object has a radius of gyration large compared with the radius $R_{f}$.
A natural extension is to consider the case where the supported load
is essentially a point mass. We can use similar methods as described
above but follow the calculation through with $\tau=0$ instead of
$F=0$. 

The stiffness of the flexure can be defined as
\begin{equation}
K_{t}=\frac{F}{y_{0}}=\alpha W\frac{\cosh\alpha L}{\alpha L\cosh\alpha L-\sinh\alpha L}.\label{eq:14}
\end{equation}
It should be noted that this expression states that the stiffness
of the flexure against transverse forces becomes infinite as the flexure
length tends to zero. This is to be expected. We can also find the
length of a simple pendulum,
\begin{equation}
R_{f}=\frac{y_{0}}{\theta_{0}}=\frac{1}{\alpha}\frac{\alpha L\cosh\alpha L-\sinh\alpha L}{\cosh\alpha L-1}.\label{eq:15-1}
\end{equation}
It is interesting to note that $R_{f}$ tends to $L-1/\alpha$ as
$\alpha L$ tends to infinity. This is consistent with the work of
Lorenzini et al \cite{Lorenzini}. The elastic component of the stiffness
is 
\begin{equation}
K_{el}=\frac{\alpha W}{4}\cdot\frac{\sinh2\alpha L-2\alpha L}{(\alpha L\cosh\alpha L-\sinh\alpha L)^{2}}.\label{eq:15}
\end{equation}
The gravitational stiffness can be written
\begin{equation}
K_{g}=\frac{\alpha W}{4}\cdot\frac{\left(2\alpha L\left(\cosh2\alpha L+2\right)-3\sinh2\alpha L\right)}{(\alpha L\cosh\alpha L-\sinh\alpha L)^{2}}.\label{eq:16}
\end{equation}
The frequency dependent component of the elastic stiffness can be
calculated straightforwardly by differentiation of the $K_{el}$ however
the resulting expression is unwieldy so we will write out the in the
case of $\alpha L\approx1$ as
\begin{equation}
K_{anel}=i\frac{\delta E^{i}}{dE}\alpha W\left(2.995-9.00\left(\alpha L-1\right)+O\left(\left(\alpha L-1\right)^{2}\right)\right).\label{eq:17}
\end{equation}
The surprising thing here is that the leading term is not unity for
this case. However this appears to be the result. Equation \ref{eq:17}
indicates that the expansion is not reliable for values of $\alpha L$
that differ too much from unity. Finally, in the case that the $L\gg\alpha^{-1}$,
we find that $K_{anel}$ tends to zero. So in the case of point mass
load the damping term tends to zero at the length of the flexure increases.
This may seem unphysical as did the previous incorrect result for
the case of the pure applied torque when we used the incorrect form
of the gravitational potential energy. Clearly as the length of the
flexure increases the energy stored in the flexure due to a finite
transverse displacement tends to zero. However any finite load must
have a finite moment of inertia and therefore there will always be
some damping due to the applied torque. At this point the quality
factor of the suspension could be calculated as a ratio of the dissipated
to stored energy for the different bending modes of the flexure. This
straighforward but is beyond the scope of of this paper. These results
agree with the paper of Cagnoli et al (2000)\cite{cagnoli}.

\section*{Conclusions}

This paper has attempted to clarify some results that were aimed at
understanding the quasi-static behaviour of flexure elements in their
role as low-loss elastic pivots for supporting beam-balances. I have
repeated the previous calculations that were published some 20 to
30 years ago and replaced them, with hindsight, with more accurate
results (that are now hopefully correct!). But note that the correct
calculation was previously published in an appendix in reference \cite{Quinn_et_al_1995}.
I have also extended the work to cover the case where the object suspended
from the flexure has a negligible radius of gyration. This may be
of use in understanding the damping of simple pendulum suspensions.
The behaviour of the predicted damping as a function of the length
of the flexure now appears to be physically reasonable in both cases.

It is worth pointing out that the above methods can be easily adapted
to inverted pendulums by switching the direction of gravity in the
problem and finding the solutions in terms of $sine$ and $cosine$
functions. 

\section*{Acknowledgements}

I am grateful to colleagues at BIPM with whom I had the pleasure of
working whilst writing the early papers. I thank Jan Harms and Conor
Mow-Lowry for taking an interest in this work. I am also grateful
to Zhang Tianxiang of Huazhong University of Science and Technology
for prompting me to write this note and also to Giuseppe Ruoso of
INFN Legnaro for useful discussions. I would like to thank the referees
for their very helpful and constructive comments on the two drafts
of this paper.

\bibliographystyle{psuthesis}
\addcontentsline{toc}{section}{\refname}\bibliography{savedrecs-2}

\end{document}